\documentclass{article}
\usepackage{booktabs}
\usepackage{graphics}
\usepackage{graphicx}
\usepackage{listings}
\usepackage{nameref}
\usepackage{algorithm}
\usepackage{algpseudocode}
\algrenewcommand\algorithmicrequire{\textbf{Input:}}
\algrenewcommand\algorithmicensure{\textbf{Output:}}
\usepackage{bm}
\usepackage{amsmath}
\usepackage{amssymb}
\usepackage{caption} 
\usepackage{multirow}
\usepackage{upgreek}
\usepackage{color}
\usepackage[table]{xcolor}

\usepackage{arxiv}
\usepackage[utf8]{inputenc} 
\usepackage[T1]{fontenc}    
\usepackage{hyperref}       
\usepackage{url}            
\usepackage{booktabs}       
\usepackage{amsfonts}       
\usepackage{nicefrac}       
\usepackage{microtype}      
\usepackage{lipsum}
\usepackage{graphicx}
\graphicspath{ {./images/} }
\begin{document}
\title{PhageMind: Generalized Strain-level Phage Host Range Prediction via Meta-learning}
\author{
   Yang Shen \\
  Dept. of Electrical Engineering\\
  City University of Hong Kong\\
  Kowloon, Hong Kong SAR, China\\
  \And
   Keming Shi \\
  Sch. of Oceanography and Earth Sciences\\
  Tongji University\\
  Shanghai, China\\
  \And
   Chen Yu \\
  College of Ocean and Earth Sciences\\
  Xiamen University\\
  Xiamen, Fujian, China\\
  \And
   Rui Zhang \\
  Institute for Advanced Study\\
  Shenzhen University\\
  Shenzhen, Guangdong, China\\
  \And
   Yanni Sun \\
  Dept. of Electrical Engineering\\
  City University of Hong Kong\\
  Kowloon, Hong Kong SAR, China\\
  \And
   Jiayu Shang \\
  Dept. of Information Engineering\\
  Chinese University of Hong Kong\\
  New Territory, Hong Kong SAR, China\\
}
\maketitle
\begin{abstract}
\textbf{Motivation:} Bacteriophages (phages) are key regulators of bacterial populations and hold great promise for applications such as phage therapy, biocontrol, and industrial fermentation. The success of these applications depends on accurately determining phage host range, which is often specific at the strain level rather than the species level. However, existing computational approaches face major limitations: many rely on genus-specific features that do not generalize across taxa, while others require large amounts of training data that are unavailable for most bacterial lineages. These challenges create a critical need for methods that can accurately predict strain-level phage-host interactions across diverse bacterial genera, particularly under data-limited conditions.

\textbf{Results:} We present PhageMind, a learning framework designed to address this challenge by enabling efficient transfer of knowledge across bacterial genera. PhageMind is trained to identify shared principles of phage-bacterium interactions from well-studied systems and to rapidly adapt these principles to new genera using only a small number of known interactions. To reflect the biological basis of infection, we represent phage-host relationships using a knowledge graph that explicitly incorporates phage tail fiber proteins and bacterial O-antigen biosynthesis gene clusters, and we use this representation to guide interaction prediction. Across four bacterial genera (\textit{Escherichia}, \textit{Klebsiella}, \textit{Vibrio}, and \textit{Alteromonas}), PhageMind achieves high prediction accuracy and shows strong adaptability to new lineages. In particular, in leave-one-genus-out evaluations, the model maintains robust performance when only limited reference data are available, demonstrating its potential as a scalable and practical tool for studying phage-host interactions across the global phageome.

\textbf{Availability:} The source code of PhageMind is available via: \href{https://github.com/YangSH-ac/PhageMind}{https://github.com/YangSH-ac/PhageMind}.

\textbf{Contact:} \href{yannisun@cityu.edu.hk}{yannisun@cityu.edu.hk}, \href{jiayushang@cuhk.edu.hk}{jiayushang@cuhk.edu.hk}
\end{abstract}
\section{Introduction}
Bacteriophages (phages) represent the most abundant and ubiquitous biological entities in the biosphere, inhabiting diverse environments ranging from the human microbiome to deep-ocean ecosystems \cite{essd-13-1251-2021}. Through their two primary life cycles (lytic and lysogenic), phages regulate bacterial abundance \cite{abu}, drive microbial evolution \cite{evo}, and modulate community structure \cite{str}. This capacity to control bacterial populations underpins the growing utility of phages in biotechnology and medicine. In the context of the global antimicrobial resistance crisis, phage therapy \cite{phathe} has emerged as a promising alternative for targeting specific pathogens. Beyond clinical applications, phages are increasingly deployed in biocontrol strategies to ensure food safety \cite{food} and to mitigate disease in aquaculture, such as Vibrio infections in farmed shrimp \cite{shrimp}. Conversely, in industrial fermentation, phage contamination of production strains poses significant economic risks, necessitating rigorous monitoring \cite{conta}. However, realizing the full potential of these ecological and biotechnological applications relies fundamentally on a precise understanding of the phage host range \cite{app}.
\label{sec:intro}
Phage-host interactions are governed by complex, multistage molecular events, while the primary determinant of host specificity is the initial adsorption step \cite{initial}. This recognition involves specific physicochemical interactions between phage receptor-binding proteins (RBPs), typically tail fibers or spikes, and bacterial surface receptors, such as lipopolysaccharides, outer membrane proteins, or capsules \cite{rbp1,lps1}. Tail fibers act as molecular sensors; their distal domains probe the bacterial surface to engage specific motifs, acting as a checkpoint for infection \cite{cp}. While subsequent steps, including genome translocation and replication, involve additional host factors, the initial binding event represents a critical filter for host susceptibility \cite{bind}. Consequently, dissecting the molecular determinants of this adsorption step offers direct insight into host specificity.

With the aid of machine learning and deep learning algorithms, many existing models can leverage the large amount of sequencing data for learning and predicting phage-host interactions at the genus \cite{g2,g3,g1} or species \cite{s2,s1,s3} resolution. However, current frameworks face substantial limitations at strain-level host prediction \cite{progress}. Phage susceptibility is frequently strain-specific; distinct strains within a single species, such as E. coli, often exhibit divergent resistance profiles due to variations in surface receptors or defense mechanisms \cite{var}. Species-level predictions lack the granularity required for precision applications, such as selecting therapeutic phages that target a pathogen while sparing commensal strains. Although recent studies have begun to address this by developing strain-level models, these efforts have generally been restricted to specific bacterial genera, lacking broader applicability.

In this work, we present a strain-level phage host range prediction framework capable of generalizing across multiple bacterial genera. By leveraging host-range data from well-characterized systems, our model utilizes a meta-learning framework \cite{metalearn} to identify recognition patterns and predict interactions in genera where experimental data are sparse. This cross-genus generalization facilitates the identification of candidate phages for specific bacterial strains, thereby accelerating the discovery of targeted phages for therapeutic, diagnostic, and ecological applications.
\subsection{Related work}
\label{sec:relate}
Recent advances in strain-level phage host range prediction have been predominantly driven by studies focusing on two model genera: \textit{Escherichia} and \textit{Klebsiella}. For instance, predictive frameworks for \textit{Escherichia} have successfully incorporated two distinct categories of host factors \cite{Escherichia}: adsorption markers (e.g., O-antigen serotypes and receptor structural variations) and defense system families (e.g., CRISPR-Cas and restriction-modification systems). This approach validates the premise that modeling specific biological barriers is essential for distinguishing strain-level susceptibility. Similarly, research on \textit{Klebsiella} has elucidated the central role of the capsular polysaccharide (CPS), or K-antigen \cite{kanti}, as the primary recognition target for phage receptor-binding proteins (RBPs). The \textit{PhageHostLearn} \cite{Klebsiella} model leverages this biological insight by employing the ESM-2 \cite{esm2} protein language model to transform RBP sequences into high-dimensional numerical vectors. By capturing the physicochemical nuances of RBP-CPS interactions, this method highlights the utility of deep learning embeddings in decoding specific recognition events.

However, relying on biological priors specific to certain genera restricts the ability of these frameworks to adapt to broader bacterial taxa. For instance, the performance of \textit{Escherichia} models depends on mature serotyping schemes and extensive experimental metadata—resources that are usually unavailable for the majority of bacterial genera. Furthermore, utilizing defense markers such as CRISPR-Cas has distinct limitations. As reported in a large-scale benchmark experiment, less than 5\% of phage-host interactions can be identified at the strain-level using CRISPR spacers matches \cite{progress}. Furthermore, CRISPR spacers indicate historical infection and they do not guarantee current susceptibility, as phages may have evolved escape mechanisms. Additionally, these approaches often require training a separate model for each individual phage genome, which is computationally expensive and difficult to scale to newly discovered phages. For example, the authors had to construct 97 distinct models to account for 97 phages \cite{Escherichia}. As the number of phages grows, the computational burden increases proportionally, rendering this approach impractical for large-scale or continuously expanding datasets. Similarly, while focusing on capsular recognition is effective for \textit{Klebsiella}, it creates a bottleneck for cross-genus application. Because capsules are not a universal bacterial trait \cite{capsule}, relying on them makes the model inapplicable to non-capsulated lineages. Thus, the field faces a trade-off: current tools offer high precision within well-characterized genera but fail to generalize to other organisms due to a lack of training data or reliance on overly specific features. This limitation highlights the critical need for a unified, taxonomy-agnostic framework capable of learning generalized recognition patterns across diverse bacterial genera.
\subsection{Overview}
\begin{figure*}[htbp]
\vspace{-0.2cm}
    \centering
    \includegraphics[width=1\linewidth]{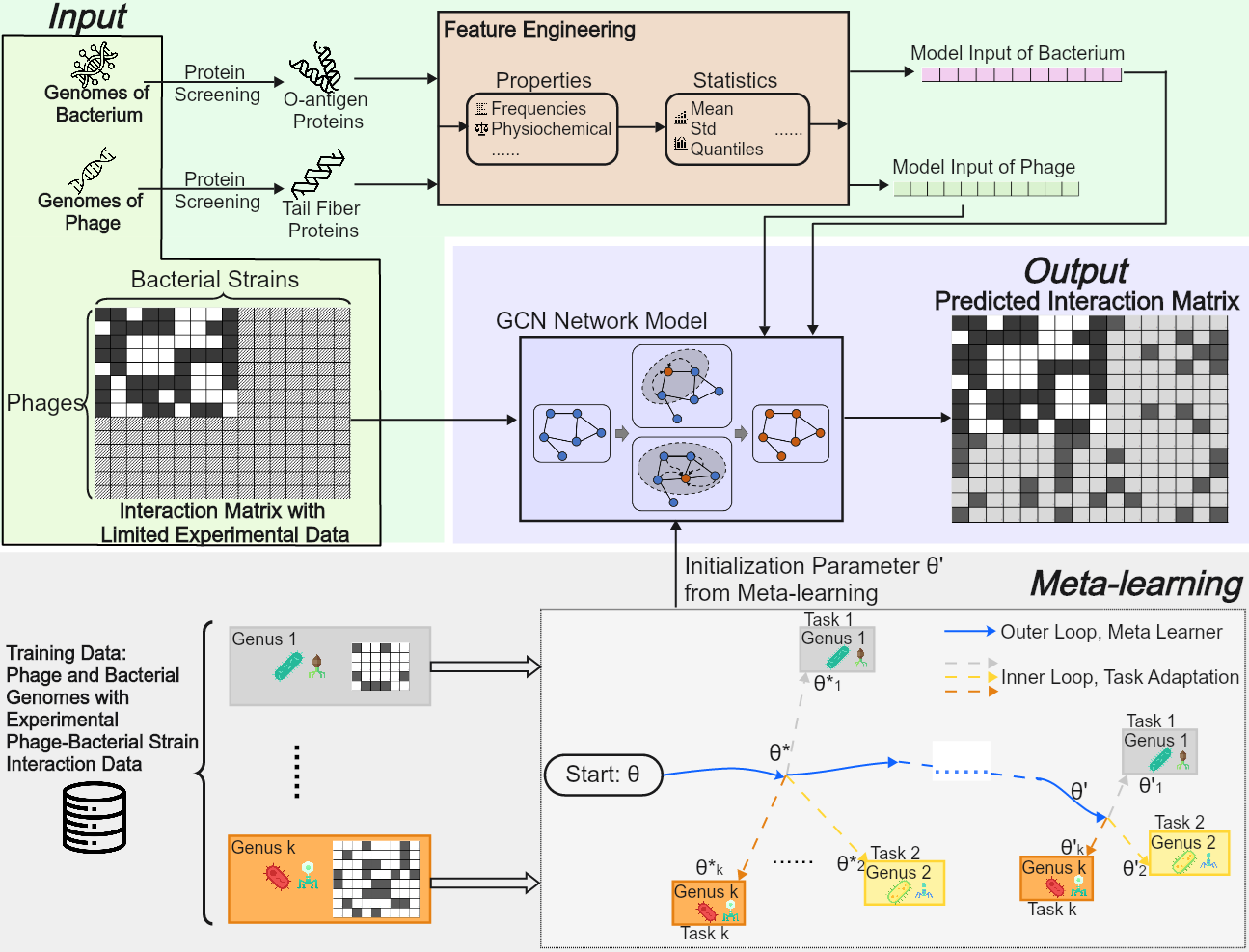}
    \vspace{-0.5cm}
    \caption {The Framework of PhageMind. It is designed to predict interactions between phages and bacterial strains within target genera, and is capable of adapting to new genus with limited data. Genomes are translated into proteins, screened, and encoded into node features using sequence‑based properties and statistical summaries, while known interactions provide edge priors for the graph convolutional network. To enhance generalization, we apply meta‑learning, which optimizes the starting parameter set $\theta$ across tasks. The meta‑learned $\theta'$ serves as a robust initialization, and can rapidly adapt to different genera, enabling improved performance across diverse genera.}
    \label{fig:pipeline} 
\end{figure*}
We present PhageMind, a domain adaptation framework for strain-level phage host range prediction (Fig. \ref{fig:pipeline}). Specifically, PhageMind is designed to predict interactions between phages and bacterial strains within target genera, and is capable of adapting rapidly to another genus even when experimental reference data are limited. To address issues that current models rely on genus-specific traits or require massive training datasets, we made two major contributions. First, we adopt biological features that are not only more relevant to phage binding but also more universal across multiple bacterial genera. Second, we integrate meta-learning \cite{metalearn} so that PhageMind distills more universal phage-bacterium attachment rules from heterogeneous multi-source data. To ensure applicability across genera, we developed a feature-encoding scheme based on conserved biosynthetic principles. We focus specifically on the O-antigen biosynthetic pathway. Because this mechanism is shared among the majority of Gram-negative bacteria, it provides a consistent framework for analyzing surface receptors. This approach allows the model to identify gene patterns associated with receptor synthesis without being constrained by the lack of established serotyping schemes in non-model organisms. Next, we incorporate meta-learning, a technique successfully applied to domain transfer and few-shot learning problems where models must generalize to new tasks with limited data. In this work, we treat experimental phage-bacterium interaction datasets from different bacterial genera as distinct domains. By learning an optimized model initialization across these domains, PhageMind can efficiently adapt to a target bacterial genus using only a small number of genus-specific samples. This strategy leads to faster convergence and improved prediction accuracy compared to conventional training methods. We validated PhageMind through a series of experiments ranging from single-genus baselines to complex cross-genus transfer tasks. The results demonstrate that our framework retains high precision within well-characterized genera while achieving robust generalization to diverse, understudied bacterial lineages with limited training data. Notably, in cross-genus adaptation experiments, PhageMind achieves performance comparable to, and in some cases exceeding that of, benchmark models even when using only 10\% of the training data compared to their full datasets. These findings establish PhageMind as a scalable solution for exploring phage-host interactions across a broad spectrum of bacteria.
\section{Methods and materials}
\label{sec:method}
We formulate strain-level phage host range prediction as a meta-learning \cite{metalearn} task to address the fundamental challenge of generalization across diverse and often data-scarce bacterial genera. Meta-learning, or ``learning to learn'', is a paradigm designed to acquire a generalizable initialization from a distribution of tasks, enabling rapid adaptation to novel tasks with minimal data. Unlike traditional supervised learning, which seeks a single decision boundary for a specific bacterial genus, meta-learning focuses on the adaptation process itself. This approach is particularly well-suited to phage-host interaction prediction, where the biological ``task'' effectively changes with the switch to a new bacterial genus. Thus, it is expected that this strategy will allow the model to capture universal interaction mechanisms while retaining the flexibility to adjust to the specific genomic variations of novel bacterial strains.

In the following sections, we first describe the curation of the strain-level interaction datasets used in this framework. Next, we introduce the workflow of the meta-learning framework, detailing the task construction and bilevel optimization strategy. We then define the base learner, explaining how biological features, specifically phage tail fibers and bacterial O-antigen clusters, are extracted and processed via a Graph Convolutional Network (GCN). Finally, we outline the training protocols, data partitioning strategies, and evaluation metrics used to validate the model.
\subsection{Strain-level phage-host interaction data}
Based on available strain-level experimental data, we constructed four genus-specific datasets to train/test our model (Table \ref{tab:data}). For the classic model genera \textit{Escherichia} and \textit{Klebsiella}, strain-level host interactions are relatively well-studied, allowing us to collect extensive interaction records from existing databases \cite{Escherichia, Klebsiella}. However, high-quality strain-level interaction data for non-model organisms is difficult to obtain. To address this gap, we collected unique experimental validation data for \textit{Vibrio} and \textit{Alteromonas}. These datasets were generated through the wet-lab experiments of our collaborators Zhang et al., involving rigorous purification, culturing, and induced lytic procedures \cite{zhang2025isolation, yu2025novel}. In each dataset, phages were carefully selected based on their potential to infect the corresponding bacterial genus, and the resulting host-phage interactions were experimentally confirmed. Importantly, the phages included in the four datasets are mutually exclusive, ensuring that no phage is shared across genera. The inclusion of this experimental data provides a valuable and rigorous benchmark for validating our framework on less characterized organisms.
\begin{table}[htbp]
\centering
\vspace{-0.3cm}
\caption{Summary of genus-specific datasets used for meta-learning tasks. The variation in interaction rates and sample sizes challenges the model to handle sparse and imbalanced distributions.}
\begin{tabular}{lrrrr}
\hline 
Dataset (Genus) & Bacteria & Phages & Exp. Pairs & Interaction Rate \\ \hline
\textit{Escherichia} & 403 & 97 & 38,435 & 20.8\% (7,976) \\
\textit{Klebsiella} & 200 & 105 & 10,006 & 3.3\% (333)\\
\textit{Vibrio} & 28 & 22 & 616 & 23.0\% (142) \\
\textit{Alteromonas} & 175 & 9 & 1,575 & 45.3\% (711)\\ \hline
\end{tabular}
\vspace{-0.2cm}
\label{tab:data}
\end{table}

As shown in Table \ref{tab:data}, these datasets vary widely in scale and interaction density. The number of bacterial strains ranges from 29 to 403, and interaction rates fluctuate between 3.3\% and 45.3\%. To faithfully reflect this natural resource imbalance, we preserve the original data distribution for each task. This compels the model to confront the central challenge of real-world biological scenarios: discovering robust patterns from inputs that are scarce, skewed, and heterogeneous.

\begin{figure}[htbp]
\vspace{-0.2cm}
    \centering
    \includegraphics[width=0.6\linewidth]{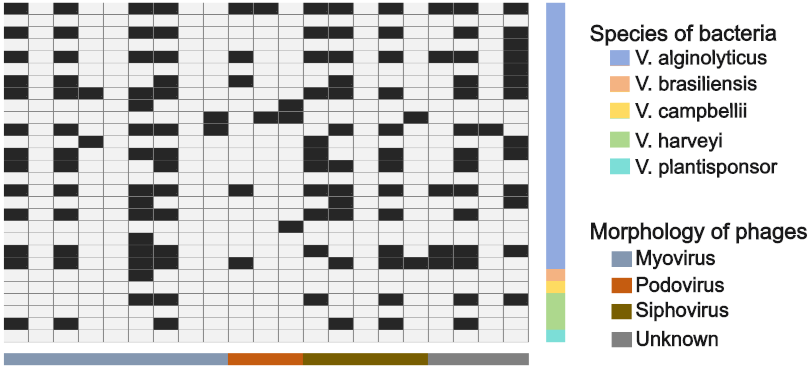}
    \vspace{-0.2cm}
    \caption{Heatmap of \textit{Vibrio} phage-host interactions (black = interaction; white = no interaction). Rows correspond to individual phages and columns to bacterial strains. Right color bar represents bacterial species while bottom color bar represents phage morphology. The map shows wide variation in phage host breadth and pronounced intra‑species differences in bacterial susceptibility, with some strains targeted by many phages while others are rarely or never infected.}
    \vspace{-0.5cm}
    \label{fig:heatmap}    
\end{figure}

To visualize the interaction patterns, we generated a heatmap for the Vibrio dataset (Fig. \ref{fig:heatmap}). The analysis reveals substantial heterogeneity within the network. Phages with identical morphology demonstrate a broad range of host breadth, with interaction counts varying from zero to nearly half of the tested bacterial strains. Even within a single morphological class, interaction profiles are distinct; some phages exhibit high overlap in their host targets, while others differ significantly. On the bacterial side, the variation is even more pronounced. Strains belonging to the same species display diverse susceptibility patterns; some strains are targeted by numerous phages, whereas others remain resistant to most or all tested phages. This uneven distribution highlights that species-level taxonomy is insufficient to capture variability in susceptibility, necessitating strain-level resolution for accurate prediction. These patterns of sparsity and heterogeneity were observed consistently across the \textit{Escherichia}, \textit{Klebsiella}, and \textit{Alteromonas} datasets, suggesting that complex, non-uniform interaction profiles are a fundamental characteristic of these phage-host networks.
\subsection{The meta-learning framework}
The implementation of PhageMind utilizes a Model-Agnostic Meta-Learning (MAML) \cite{maml} strategy designed to explicitly address the challenge of cross-genera generalization. Given the vast diversity of phage-host interaction patterns, it is critical to prevent the model from overfitting to genus-specific features. We employ a bilevel optimization process that mimics the scenario of encountering a novel genus: an ``inner loop'' performs rapid adaptation using a small support set of samples, while an ``outer loop'' optimizes the model’s initialization based on performance across multiple distinct genera. This episodic training regime encourages the learning of highly transferable feature representations and weight initializations. Consequently, when deployed on an new genus, PhageMind can rapidly adapt to establish accurate decision boundaries using only a few samples, effectively overcoming the data scarcity limitations inherent in traditional deep learning approaches. 
\subsubsection{Task construction and data organization}
The foundation of our framework is the organization of biological data into distinct ``tasks''. In this context, we define a single ``task'' as the prediction of phage host range within a specific bacterial genus. This design not only reflects the biological reality that interaction mechanisms (such as receptor specificity) often vary significantly between genera, but also aligns with the practical needs of laboratories and tool users who are typically focused on specific bacteria relevant to their own research.
\subsubsection{Optimization flow: the inner and outer loops}
\begin{figure}[htbp]
    \centering
    \vspace{-0.2cm}
    \includegraphics[width=0.7\linewidth]{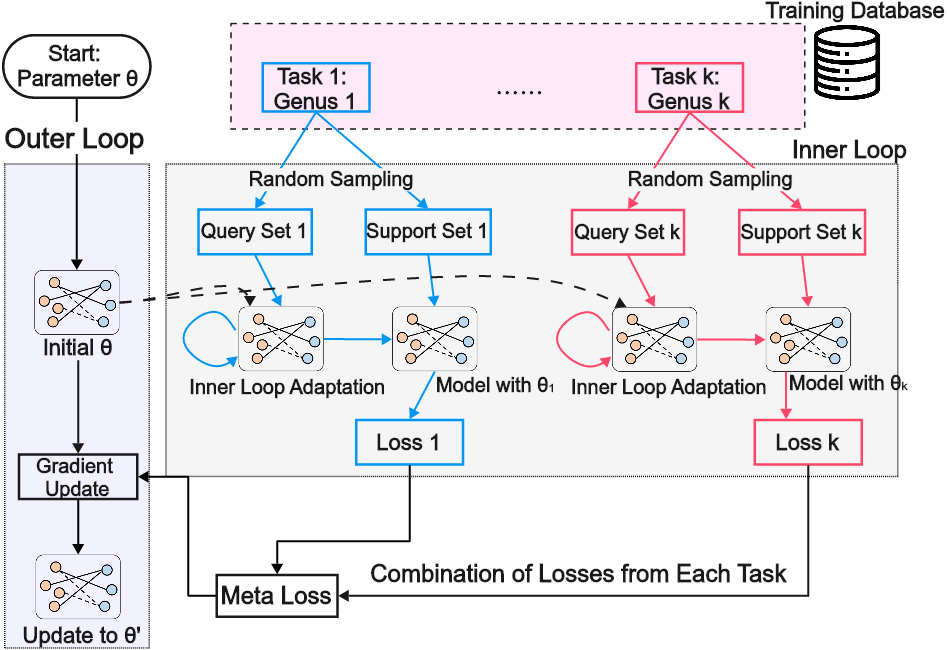}
    \vspace{-0.2cm}
    \caption{Structure of our meta-learning framework. Each independent task corresponds to a mutually exclusive dataset representing a single genus. Within each task, training data is split into a support set for inner-loop adaptation and a disjoint query set for evaluation. The outer loop performs a meta-update by aggregating losses across tasks to optimize meta-parameters.}
    \vspace{-0.5cm}
    \label{fig:ml}    
\end{figure}

The core logic of our meta-learning framework involves a dual-loop optimization process (Fig. \ref{fig:ml}). This structure allows the model to act as a ``generalist'' in the outer loop (learning broad interaction rules) while temporarily becoming a ``specialist'' in the inner loop (adapting to a specific genus). For every sampled task (genus), the training data is split into two mutually exclusive subsets:
\begin{itemize}
    \item \textbf{Support Set (Inner Loop):} A small set of labeled interaction pairs used for rapid, temporary adaptation (30\% of the training set).
    \item \textbf{Query Set (Outer Loop):} A set of non-overlapping interaction pairs used to evaluate generalization (70\% of the training set).
\end{itemize}
\paragraph{The inner loop (the specialist)}
The goal of the inner loop is to simulate the process of fine-tuning the model to a specific task. Starting from the model's current global parameters ($\theta$), the framework performs a few steps of gradient descent (typically five iterations) using only the support set. This produces a set of temporary, task-specific parameters ($\theta_i$). The process requires only seconds of computation but allows the network to capture the distinctive patterns of that specific genus (e.g., specific O-antigen receptor structures).
\paragraph{The outer loop (the generalist)}
The outer loop evaluates the quality of the ``specialist'' created in the inner loop. We test the model on the query set, using the temporary parameters $\theta_i$. The loss calculated here represents how well the model generalized to unseen data within that genus. Crucially, the meta-gradient update is computed based on this query loss to update the global parameters to $\theta'$. By minimizing the loss on the query set, the model effectively learns an initialization state that is easy to fine-tune, ensuring rapid convergence even on data-scarce genera.
\subsection{The base model: graph convolutional network}
While the meta-learning framework dictates how our framework learns, the GCN serves as the core learner that processes the data. In our framework, GCN explicitly encodes the topological relationships between phages and bacteria by iteratively aggregating embedding information from neighbors in the Knowledge graph, allowing it to uncover universal interaction patterns.
\subsubsection{Knowledge graph construction}
We model the phage-host interaction landscape as a bipartite graph $G = (B, P, E)$, where $B$ represents the set of bacterial nodes and $P$ represents the set of phage nodes. Edges $E$ are constructed strictly based on experimentally validated infection relationships. To further strengthen the model’s robustness, particularly for predicting interactions in new genera, we introduce a dynamic edge masking strategy. During training, we apply a masking rate of 0.7, randomly hiding a portion of known infection edges during training. This compels the model to infer potential connections primarily from the intrinsic molecular features of the nodes and the broader structural context, rather than memorizing direct links.
\subsubsection{Feature Engineering and Embedding}
The biological mechanism of phage infection is fundamentally a molecular recognition event, often characterized as a ``lock-and-key'' interaction. In contrast to alignment-free methods that rely on whole-genome information, which often introduce noise from non-interacting genomic regions, our framework explicitly targets the specific molecular determinants that govern the physical interface between the phage and the host.
\paragraph{Phage Side: Tail Fiber Proteins}
For bacteriophages, the primary determinants of host range are the tail fiber proteins (also known as receptor-binding proteins). These structures function as the recognition machinery of the phage, specifically binding to receptors on the bacterial surface. Consequently, rather than embedding the entire phage genome, we isolate tail fiber protein sequences to capture the specific ``key'' mediating infection. To ensure high-quality input, we implemented a multi-stage identification pipeline that integrates automated annotation (Pharokka \cite{pharokka}) with structural validation (AlphaFold2 \cite{alphafold2}), thereby verifying the domain architecture of these critical proteins.
\paragraph{Bacterial Side: O-antigen Biosynthesis Clusters}
On the host side, the corresponding ``lock'' is the O-antigen component of the lipopolysaccharide, which forms the outermost layer of Gram-negative bacteria. As a highly variable structure, the O-antigen serves as the primary receptor for numerous phages. Accordingly, we extract O-antigen biosynthesis gene clusters to represent the core bacterial feature. Given that these clusters exhibit significant variation in both length (5 to 18 genes) and composition across genera \cite{oa518}, we developed a robust localization strategy. By utilizing conserved housekeeping genes as genomic anchors (e.g., \textit{galF}, \textit{gnd}, \textit{hisI}), we precisely extracted the functional units responsible for surface receptor synthesis (see Supplementary Note 1 for details). This ensures the model learns from the specific surface topology encountered by the phage.
\paragraph{Node feature} To encode these biological sequences into inputs of the GCN, protein sequences and their corresponding DNA sequences were converted into numerical embeddings. We calculated features specifically relevant to protein-protein binding, such as hydrophobicity, electrostatic charge, and secondary structure elements (helix/sheet/turn) \cite{deeppbi}. These properties are the physical drivers of the binding affinity between the phage tail and the bacterial receptor. Then, the calculated features from bacteria and phages are passed through independent fully connected layers to map them into a shared latent space, eliminating scale discrepancies. The equation of this Heterogeneous FC Layer can be listed in Eqn. \ref{eq:heteromlp}.
\begin{equation}
    \label{eq:heteromlp}
    H_B=\text{ReLU}(W_B\times X_B), \hspace{0.4cm}H_P=\text{ReLU}(W_P\times X_P)
\end{equation}
\noindent where $X_B\in \mathbb{R}^{d_B\times n_B}$ and $X_P\in \mathbb{R}^{d_P\times n_P}$ denote the raw feature matrices of bacteria and phages, respectively. $n_P$ and $n_B$ are the number of phages and bacteria. $d_B$ and $d_P$ are the dimensions of the input phage and bacterial features. $W_B\in \mathbb{R}^{d\times d_B}$ and $W_P\in \mathbb{R}^{d\times d_P}$ are the weight matrices that project the inputs into the shared latent space of dimension $d$. ReLU is the element-wise rectified linear unit activation function.
\begin{figure}[htbp]
    \vspace{-0.2cm}
    \centering
    \includegraphics[width=0.75\linewidth]{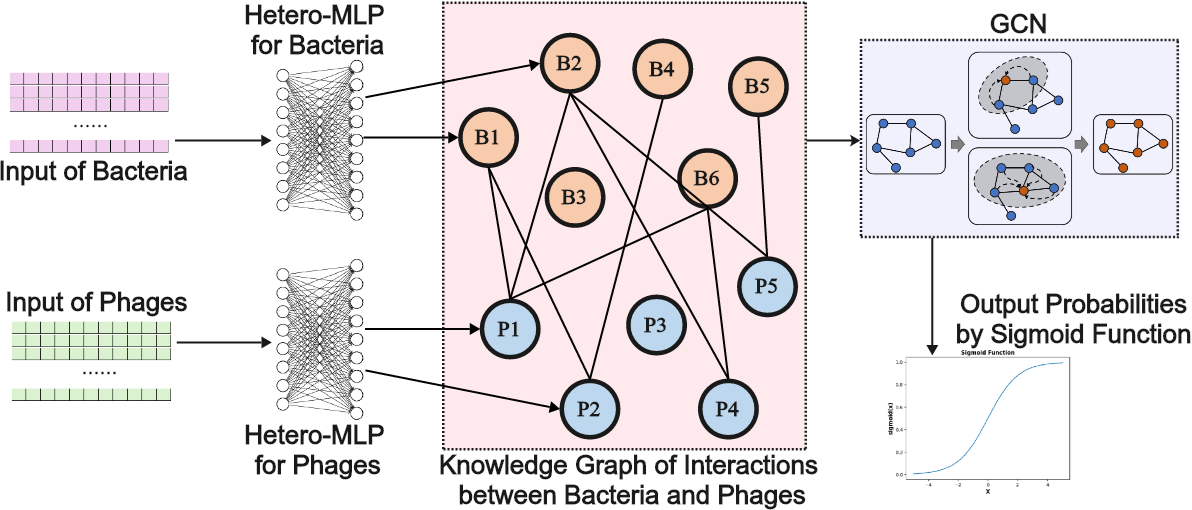}
    \vspace{-0.2cm}
    \caption{Structure of our GCN Model. Bacteria and phages nodes form mutually exclusive subsets, connected only through experimentally verified infection relationships. }
    \vspace{-0.3cm}
    \label{fig:gcn}    
\end{figure}
\subsubsection{Graph Convolution and Interaction Prediction}
We utilize a two-layer GCN architecture to unlock implicit biological information. The GCN layer can be expressed in Eqn. \ref{eq:gcn}. The first layer aggregates information from direct neighbors (explicit interactions). The second layer establishes indirect information channels between nodes that share common partners. For example, two bacteria that are not directly linked may share similar phage susceptibilities; the second GCN layer allows these nodes to ``perceive'' the topology through the shared phage node. This mechanism is particularly valuable for microbiome data, which is often sparse.
\begin{equation}
    \label{eq:gcn}
    H^{(l+1)}=\text{ReLU}(\tilde{D}^{-\frac{1}{2}}\tilde{A}\tilde{D}^{-\frac{1}{2}}H^{(l)}W^{(l)})
\end{equation}
\noindent where $\tilde{A}=A+I$, $\tilde{D}_{ii}=\sum_j\tilde{A}_{ij}$. Adjacency matrix $A\in R^{n\times n}$ encodes the connections between $n$ nodes in the graph. Degree matrix $D\in R^{n\times n}$ is a diagonal matrix whose diagonal entries are node degrees. $H^{(0)}$ is the embedding from Eqn \ref{eq:heteromlp}.

Following graph convolution, the high-level embeddings of a target bacterium and phage are fused via matrix addition. This joint representation is passed through a two-layer fully connected network which performs non-linear transformations to refine the signal. Finally, a sigmoid activation function maps the output to a probability score $P \in [0, 1]$, representing the likelihood of a successful infection.
\subsection{Model training} 
\begin{figure}[htbp]
    \vspace{-0.2cm}
    \centering
    \includegraphics[width=0.65\linewidth]{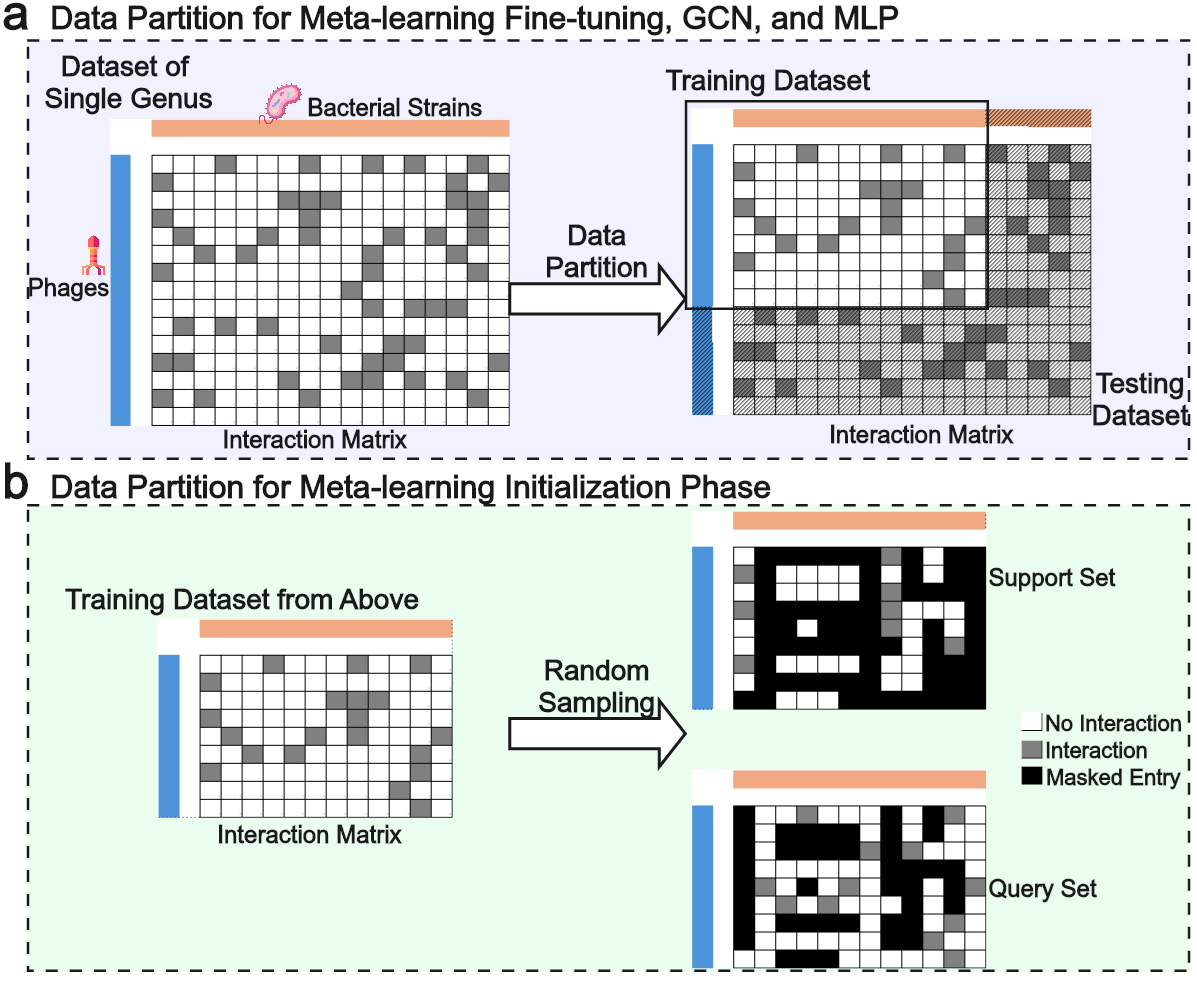}
    \vspace{-0.2cm}
    \caption{Strategy we used for dataset partition. (a) Double‑isolation is used for data partition in each bacterial genus. Bacteria and phages are independently partitioned into training and test set with ratio 7:3. The training set contains only interactions among training organisms, while the test set includes all remaining pairs (training-test, test-training, and test-test). (b) Each training set is randomly split into support set (30\%) and query set (70\%), which are needed to implement meta-learning framework.}
    \label{fig:data_part}    
    \vspace{-0.2cm}
\end{figure}

To enable rigorous evaluation and avoid data leakage, we designed a ``double isolation'' strategy for dataset partitioning (Fig. \ref{fig:data_part}a). Bacteria and phages are independently divided into training and test pools. The training phase is restricted to interactions only between training organisms, while the test set encompasses all other pairs involving unseen entities: training bacteria vs. test phages, test bacteria vs. training phages, and test bacteria vs. test phages. This strict separation forces the model to learn molecular recognition rules rather than memorizing specific strain identities, thereby faithfully simulating the exploratory nature of real-world phage-host interaction prediction.
In our experiments, the data were partitioned into training and test sets at a 7:3 ratio. Following a standard meta-learning protocol, the training data for each task were randomly divided into support (30\%) and query (70\%) sets at each iteration (epoch) to learn the initialization parameters (Fig. \ref{fig:data_part}b). To enhance model robustness, particularly for predicting interactions within novel genera, we introduced a dynamic edge masking strategy. During the training phase, we applied a masking rate of 0.7, randomly concealing a subset of known infection edges. Following the meta-learning process, we performed fine-tuning to adapt the shared initialization parameters to specific genera using their respective training sets until convergence. Final performance was evaluated on the held-out test sets to quantify the generalization capability of the PhageMind model.
\subsection{Metrics} 
In binary classification that aims to predict bacterium-phage interactions, the Area Under the ROC Curve (AUC) serves as a widely accepted metric for evaluating model performance \cite{Escherichia, Klebsiella}. Mathematically, AUC is defined as the integral of the area beneath the Receiver Operating Characteristic (ROC) curve, and its calculation follows a well-established formula as follows:
\begin{equation}
    \label{eqauc}
    \text{AUC} = \int_{0}^{1}TPR\cdot FPR\hspace{0.1cm}d(FPR)
    \vspace*{-0.1cm}
\end{equation}
where
\begin{equation}
TPR=\frac{TP}{TP+FN},\qquad FPR=\frac{FP}{FP+TN}
\end{equation}
The True Positive Rate (TPR) reflects the model’s ability to correctly identify genuine interacting pairs, while the False Positive Rate (FPR) measures the frequency of errors in which non-interacting pairs are misclassified as interacting. The ROC curve itself plots the relationship between TPR and FPR across all possible classification thresholds, and AUC provides a single quantitative value that summarizes the overall quality of this curve. The value of AUC lies within the interval [0, 1] with values closer to 1 indicating a stronger ability to distinguish between positive and negative samples. A high AUC score means that the model can, with high probability, rank a randomly selected true interacting pair ahead of a randomly selected non-interacting pair. 
\section{Result}
\label{sec:result}
To the best of our knowledge, PhageMind is the first tool capable of performing strain-level phage host range prediction across multiple bacterial genera. In contrast, existing host prediction tools are primarily designed to operate at the genus level or higher taxonomic ranks. As a result, no existing methods can be benchmarked fairly against PhageMind using the same input data and prediction resolution. Therefore, we implemented alternative models trained on the same type of training data as PhageMind. First, we implemented a multiple layer perceptron model as a standard deep learning baseline. Furthermore, as GCN has been successfully applied for phage host prediction in several methods \cite{s2, du2023prokaryotic, liu2025phpgat} and is an important component in our framework, we also implemented GCN without the meta-learning strategy as an ablation study.

The following results are presented in a progression of increasing difficulty, designed to systematically validate each component of the framework. First, we perform an ablation study to verify that our specific feature engineering and GCN architecture provide a solid foundation for capturing interaction topology (Section 3.1). Building on this foundation, we further examine how the model learns during training, demonstrating that the meta-learning approach converges more quickly and exhibits more stable training behavior than baseline models (Section 3.2). Finally, we address the most challenging and realistic scenario: cross-genus generalization. By withholding an entire genus during the meta-training stage, we evaluate the framework’s ability to adapt to new bacterial genera using only a limited number of labeled examples, simulating the challenge of predicting interactions for new bacterial strains with sparse experimental data (Section 3.3).
\subsection{Explicit encoding of molecular interfaces and interaction topology improves prediction accuracy}
\begin{figure}[htbp]
    \vspace{-0.2cm}
    \centering
    \includegraphics[width=0.6\linewidth]{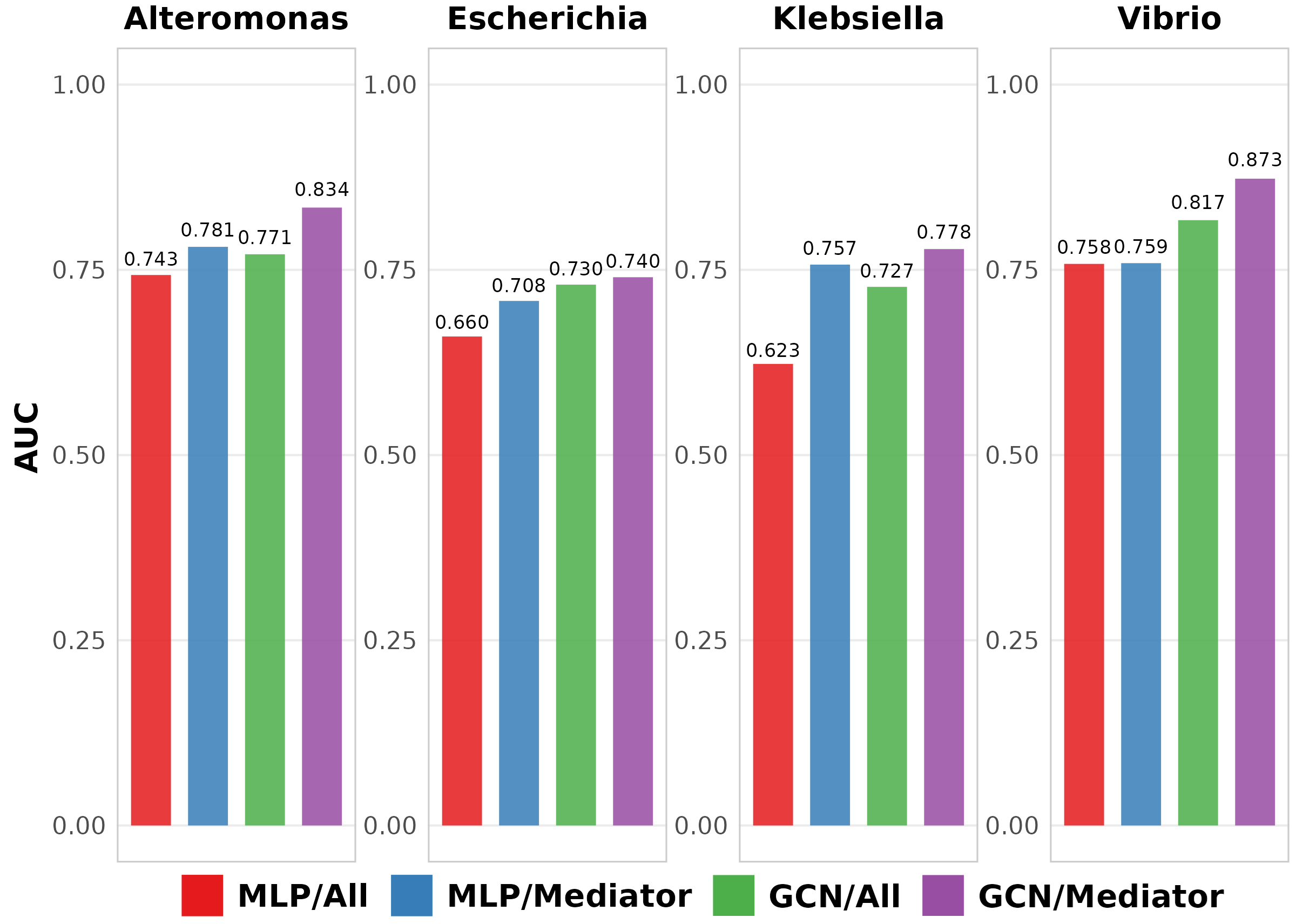}
    \vspace{-0.2cm}
    \caption{AUC barplot for four genera by using all proteins or only our chosen features (mediators).  Mediator: using features from O-antigen proteins on the bacterial side and tail fiber proteins on the phage side; All: using features from all proteins in both side. Across all the four genera, employing GCN and using the features from mediators yields an increase in AUC.}
    \vspace{-0.5cm}
    \label{fig:roc_single}    
\end{figure}

Before addressing the adaptability of our framework, we first evaluated the biological and computational hypotheses underlying the model's design. Phage-host specificity is governed by molecular recognition events, typically between phage tail fibers and bacterial surface receptors like the O-antigen, rather than broad genomic similarity. To test whether explicitly encoding this mechanism improves predictive performance, we compared using two distinct feature sets: one restricted to known interaction mediators (tail fibers and O-antigens) and another utilizing all available proteins (Fig. \ref{fig:roc_single}). To focus on the feature comparison, we only use the GCN component or MLP in this experiment. 

Across the four genera analyzed (\textit{Escherichia}, \textit{Klebsiella}, \textit{Vibrio}, and \textit{Alteromonas}), models restricted to specific interaction proteins consistently outperformed the approaches utilizing all proteins. Notably, restricting input features to tail fibers and O-antigens improved AUC values by 3-10\% using a MLP and by 2-6\% using a GCN. Thus, by excluding non-interacting proteins, the models can prioritize the molecular interfaces governing infection, thereby reducing the risk of overfitting to irrelevant genomic features.

We next examined the necessity of modeling the interaction topology by the GCN against the MLP. The MLP processes each phage-bacteria pair in isolation, while the GCN aggregates information from neighboring nodes in the bipartite graph. The results in Fig. \ref{fig:roc_single} and supplementary Fig. S1 indicate that the topological approach provides better predictive capability. In the \textit{Klebsiella} dataset, which is the most imbalanced dataset, the MLP showed lower generalization performance, whereas the GCN maintained discriminative power. This suggests that the neighborhood aggregation mechanism of graph convolutions assists in inferring links by utilizing the known interactions of similar species. Similarly, in the \textit{Alteromonas} dataset, which has fewer samples, the GCN outperformed the MLP. The performance hierarchy across datasets, where the GCN with specific proteins ranks highest, followed by the GCN with all proteins, and finally the MLP variants, supports our structural hypothesis. By combining molecular feature selection with a network-based architecture, the base learner can effectively model the biophysics of infection, providing a stable foundation for the subsequent meta-learning tasks.
\subsection{Meta-learning enables adaptation and robust convergence}
Then, we examined whether the meta-learning framework can achieve improved performance. A major challenge in applying deep learning to biological problems is the substantial amount of data and computational effort typically required to train models from scratch. This limitation is particularly pronounced in phage-host studies, where experimentally validated interaction data are often scarce. We hypothesized that meta-learning helps overcome these challenges by learning a broadly informative starting point (initialization parameters) that allows the model to adapt efficiently to different bacterial strains.
\begin{figure}[htbp]
    \vspace{-0.2cm}
    \centering
    \includegraphics[width=0.65\linewidth]{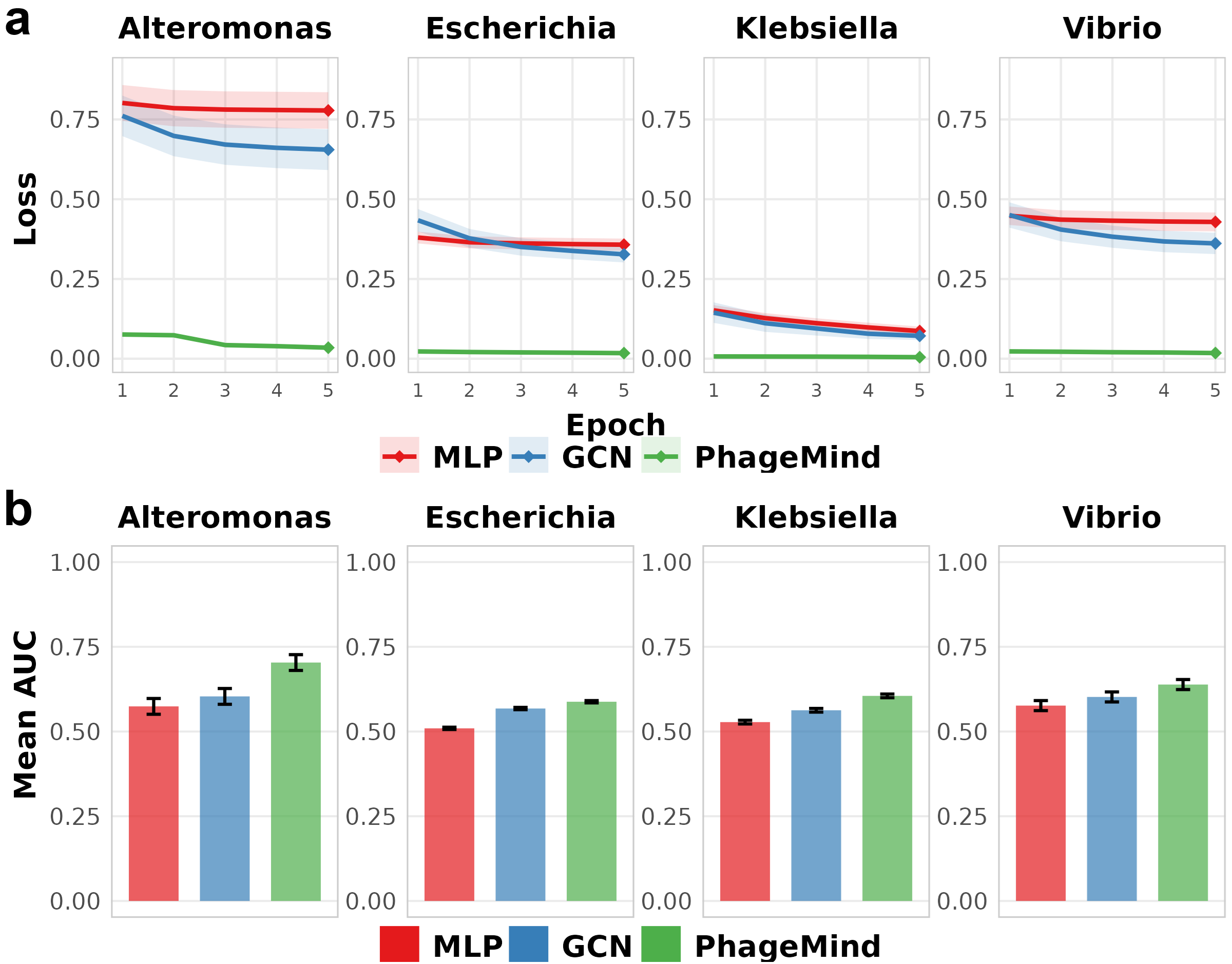}
    \vspace{-0.2cm}
    \caption{Performance of models after 5 epochs training. (a) Dynamics of early training loss (first 5 epochs). The line graph plots the average loss changes across four genera, with shaded areas representing 95\% confidence interval.  (b) Mean AUC after rapid fine-tuning to the four genera. The barplot shows the mean AUC after training for only 5 epochs with error bars representing 95\% confidence interval.}
    \vspace{-0.2cm}
    \label{fig:loss_auc}    
\end{figure}

To test this hypothesis, we first evaluated training efficiency by monitoring model loss during the early stages of adaptation (Fig. \ref{fig:loss_auc}a). Compared with a conventionally trained GCN and MLP initialized at random, the meta-learned model began training with substantially lower prediction error. This observation suggests that, during meta-training, the model had already captured general and reusable patterns of phage-bacterium interactions across multiple bacterial genera. As a result, the model did not need to rediscover these fundamental features when adapting to a new target genus. Consequently, the meta-learned model converged much more rapidly, reaching high predictive accuracy with far fewer training iterations than the baseline model. In contrast, the model trained from scratch exhibited a prolonged initial learning phase. This difference in training behavior translates into a clear advantage under data-limited conditions. When the number of training epochs was restricted (Fig. \ref{fig:loss_auc}b), the meta-learning framework achieved approximately 85-90\% of its maximum AUC within only five epochs, whereas the baseline model required nearly ten times longer to reach comparable performance.

Together, these results demonstrate that meta-learning confers a practical few-shot learning capability, enabling rapid and stable adaptation to new bacterial strains using limited training data and reduced computational resources. This property is particularly valuable for strain-level phage host prediction, where newly sequenced or poorly characterized bacterial strains are common and experimental interaction data remain limited.

While computational efficiency is important, a practical predictive framework must also be reproducible and reliable. Deep learning models trained on small or sparse biological datasets often exhibit instability, leading to inconsistent results or convergence failures. To assess model robustness, we analyzed the stability of the training process across 40 independent trials. We compared our approach against two baseline training strategies for both GCN and MLP. The first strategy involved standard genus-specific training, where models were trained and tested only on interactions within a single genus. The second strategy involved pooling training data from \textit{Alteromonas}, \textit{Escherichia}, \textit{Klebsiella}, and \textit{Vibrio} to create a combined dataset. This second strategy ensured that the baseline models had access to a sample size equivalent to that used in our meta-learning framework, allowing us to determine whether performance gains were due to the methodology or simply data availability.
\begin{figure}[htbp]
    \centering
    \vspace{-0.2cm}
    \includegraphics[width=0.65\linewidth]{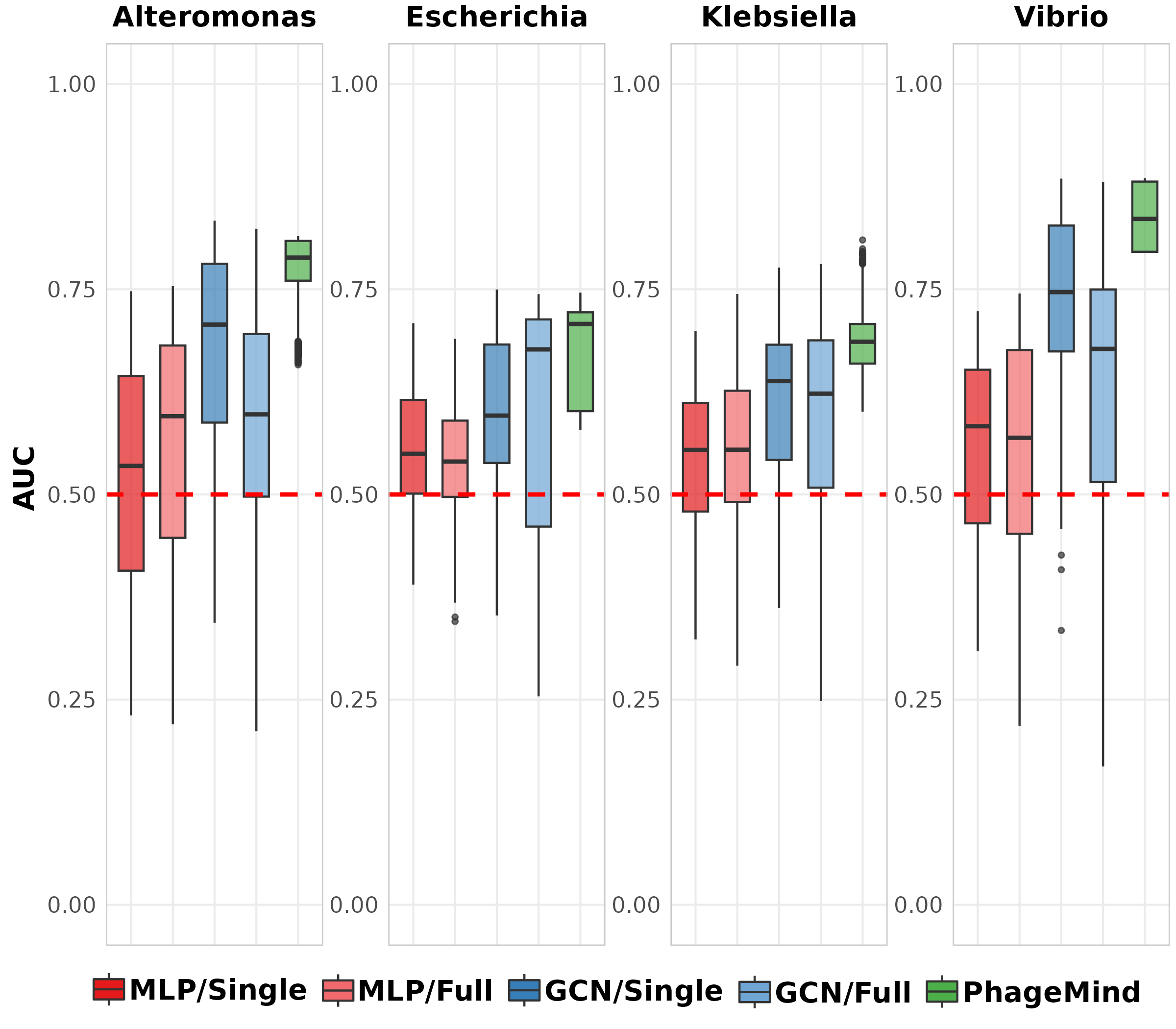}
    \vspace{-0.2cm}
    \caption{AUC comparison of PhageMind and baseline models trained from scratch (GCN and MLP). Boxplots show the AUC distributions across four bacterial genera. GCN/Single and MLP/Single refer to models trained on a single‑genus dataset, while GCN/Full and MLP/Full refer to models trained on the combined datasets from all four genera. PhageMind achieves higher AUC values compared to GCN and MLP.}
    \vspace{-0.2cm}
    \label{fig:auc_ml}    
\end{figure}

As shown in Fig. \ref{fig:auc_ml} and supplementary Fig. S2, the results of the baseline models were unstable, exhibiting wide performance variance and occasional convergence failures (AUC \textless 0.5), particularly on the complex \textit{Klebsiella} dataset. In contrast, the PhageMind framework maintained consistent performance across trials. Notably, simply increasing the volume of data for standard models proved ineffective; the GCN and MLP models yielded lower performance when trained on the combined dataset than when trained on genus-specific tasks. This indicates that PhageMind’s superior performance stems from its meta-learning strategy, specifically its optimized initialization, rather than the scale of the training data. Thus, PhageMind offers a robust solution for strain-level host range prediction.
\subsection{Cross-genus adaptation enables targeted prediction in new host lineages}
To evaluate the framework’s capacity for discovery, specifically when researchers are handling bacterial lineages absent from existing databases, we conducted a leave-one-dataset-out experiment. In this protocol, we systematically excluded an entire genus and its associated phages during the meta-training phase, using the remaining three genera to learn the initialization parameters. The excluded genus was then treated as a novel target task. We adapted the model to this ``new'' lineage using varying amounts of labeled examples settings, simulating the real-world challenge of characterizing new organisms with sparse experimental data.

\begin{figure}[htbp]
    \vspace{-0.2cm}
    \centering
    \includegraphics[width=0.65\linewidth]{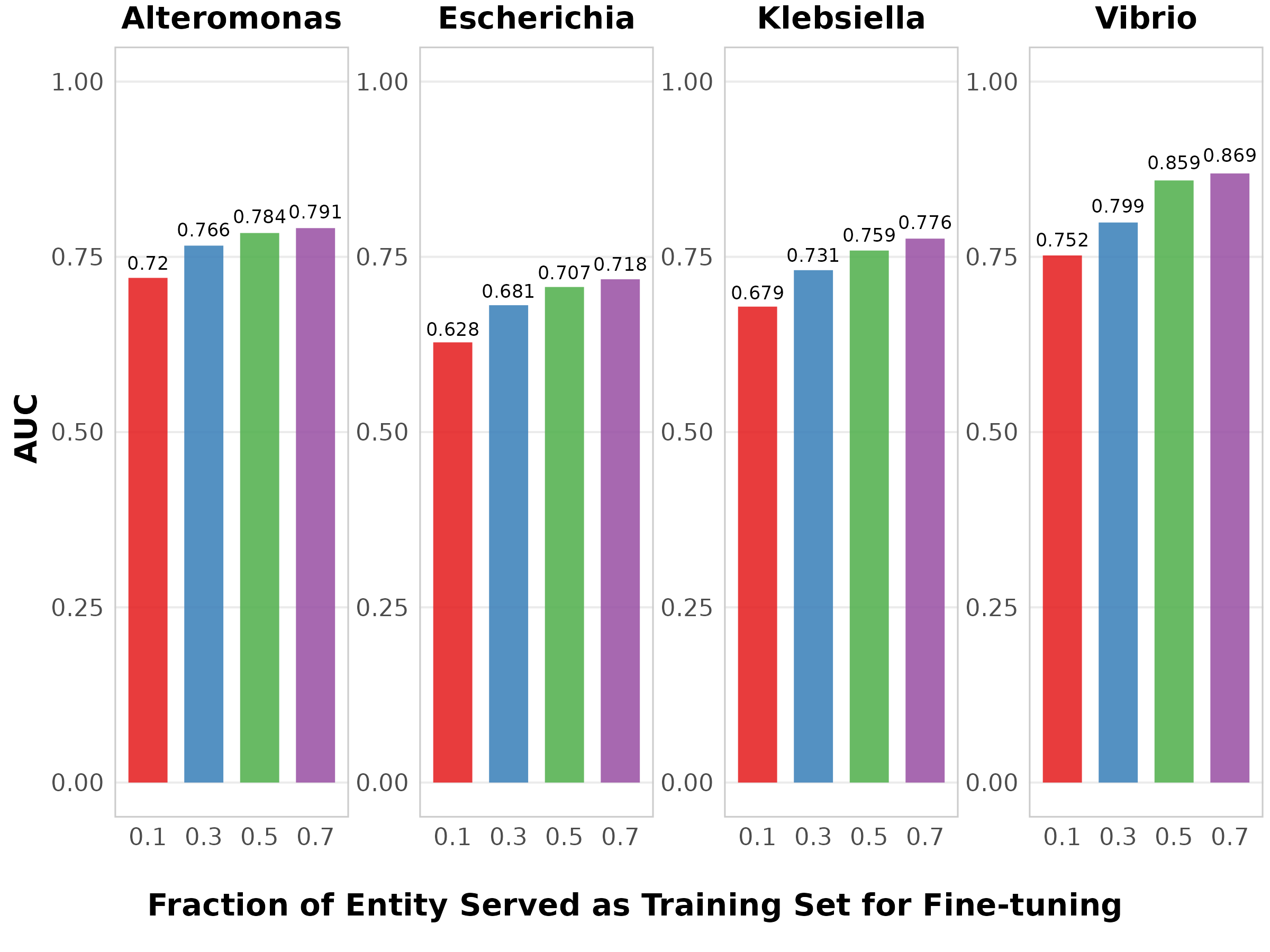}
    \vspace{-0.2cm}
    \caption{Each bar represents the AUC for a specific genus relative to the fraction of entities used for fine-tuning. While AUC gradually declines as the training fraction decreases, performance remains robust even with only 10-30\% of the data. At the 10\% threshold, sample sizes are small: \textit{Vibrio} comprises approximately 3 bacteria and 2 phages; \textit{Escherichia}, 40 bacteria and 10 phages; \textit{Klebsiella}, 20 bacteria and 10 phages; and \textit{Alteromonas}, 17 bacteria and 1 phage.}
    \vspace{-0.2cm}
    \label{fig:auc_cg}    
\end{figure}

As shown in Fig. \ref{fig:auc_cg} and supplementary Fig. S3, PhageMind demonstrated robust generalization to these ``new'' genera, maintaining high predictive accuracy even under strict data constraints. The performance remained stable even when we simulated a few-shot learning scenario by reducing the training data to fractions of 0.5, 0.3, and 0.1 (Fig. \ref{fig:auc_cg}). Remarkably, even when restricted to only 10\% of the available data, the meta-learning framework achieved performance comparable to—and in some cases exceeding—GCN and MLP models trained on the full dataset (Fig. \ref{fig:auc_ml}). This indicates that meta-learned initialization provides a highly effective "head start," allowing the model to establish reliable decision boundaries without the massive datasets typically required by traditional deep learning frameworks, thereby significantly reducing the experimental burden of mapping new host genera.
\section{Discussion}
In this study, we introduce PhageMind, a framework that integrates meta-learning with GCNs to address the challenge of strain-level phage host range prediction. By leveraging a meta-learning paradigm, PhageMind distills fundamental, cross-genera interaction patterns from different bacterial genera, establishing genus-agnostic priors. These priors function as an optimized initialization state, enabling the model to rapidly adapt to new bacterial genera using only limited data. Complementing this, the GCN component embeds the specific molecular determinants that govern the physical interface between the phage and the host within a heterogeneous graph structure, thereby enhancing molecular signals with topological insights derived from known interactions. This synergistic approach allows PhageMind to construct robust decision boundaries between phage-host pairs across domains. Our comprehensive evaluation demonstrates that this strategy not only achieves high predictive accuracy but also significantly accelerates convergence and reduces training variance compared to the baseline models.

Despite these gains, PhageMind has some limitations that point to future work. First, the GCN relies on a restricted but known set of interaction edges. When no interaction information is available for a new genus, the graph signal is weak and the meta‑learned initialization alone may yield only modest results. Second, although we identified a broadly applicable protein-cluster feature set, more precise protein clusters could be defined for different genera if detailed mechanisms are understood. Selecting genus-appropriate protein clusters would likely improve feature specificity and thus prediction accuracy. Third, our current evaluation is constrained by the nature of experimental validation. While successful induction yields high-confidence positive pairs, negatives may be false due to lab conditions failing to trigger lysis rather than true incompatibility. Expanding to more diverse genera and using rigorous experimental designs to address potential false negatives is needed to identify universal infectivity markers and improve utility for therapy and diagnostics.
\section*{Funding}
This work was supported by the Hong Kong Research Grants Council (RGC) General Research Fund (GRF) [11209823], the City University of Hong Kong [9667256, 9678241, 7005866], the National Natural Science Foundation of China (32570002), and the Shenzhen Science and Technology Program (JCYJ20241202124403006).
\bibliographystyle{unsrt}  
\bibliography{references}
\end{document}